\documentclass[preprint,12pt]{elsarticle}

\usepackage{graphicx}

\usepackage{amssymb}

\journal{Physics Letters B}

\begin{document}

\begin{frontmatter}



\title{Search for solar axions in XMASS, a large liquid-xenon detector}


\author[ICRR,IPMU]{K.~Abe}
\author[ICRR]{K.~Hieda}
\author[ICRR,IPMU]{K.~Hiraide}
\author[ICRR]{S.~Hirano}
\author[ICRR,IPMU]{Y.~Kishimoto}
\author[ICRR,IPMU]{K.~Kobayashi}
\author[ICRR,IPMU]{S.~Moriyama}
\author[ICRR]{K.~Nakagawa}
\author[ICRR,IPMU]{M.~Nakahata}
\author[ICRR,IPMU]{H.~Ogawa}
\author[ICRR]{N.~Oka}
\author[ICRR,IPMU]{H.~Sekiya}
\author[ICRR]{A.~Shinozaki}
\author[ICRR,IPMU]{Y.~Suzuki}
\author[ICRR,IPMU]{A.~Takeda}
\author[ICRR]{O.~Takachio}
\author[ICRR]{K.~Ueshima\fnref{tohoku}}
\author[ICRR]{D.~Umemoto}
\author[ICRR,IPMU]{M.~Yamashita}
\author[ICRR]{B.~S.~Yang}

\author[GIFU]{S.~Tasaka}

\author[IPMU]{J.~Liu}
\author[IPMU]{K.~Martens}

\author[KOBE]{K.~Hosokawa}
\author[KOBE]{K.~Miuchi}
\author[KOBE]{A.~Murata}
\author[KOBE]{Y.~Onishi}
\author[KOBE]{Y.~Otsuka}
\author[KOBE,IPMU]{Y.~Takeuchi}

\author[KRISS]{Y.~H.~Kim}
\author[KRISS]{K.~B.~Lee}
\author[KRISS]{M.~K.~Lee}
\author[KRISS]{J.~S.~Lee}

\author[MIYA]{Y.~Fukuda}

\author[NAGOYA,KMS]{Y.~Itow}
\author[NAGOYA]{K.~Masuda}
\author[NAGOYA]{Y.~Nishitani}
\author[NAGOYA]{H.~Takiya}
\author[NAGOYA]{H.~Uchida}

\author[SEJONG]{N.~Y.~Kim}
\author[SEJONG]{Y.~D.~Kim}

\author[TOKAI1]{F.~Kusaba}
\author[TOKAI2]{D.~Motoki\fnref{tohoku}}
\author[TOKAI1]{K.~Nishijima}

\author[YNU]{K.~Fujii}
\author[YNU]{I.~Murayama}
\author[YNU]{S.~Nakamura}

\address[ICRR]{Kamioka Observatory, Institute for Cosmic Ray Research,
  the University of Tokyo, Higashi-Mozumi, Kamioka, Hida, Gifu, 506-1205, Japan}
\address[GIFU]{Information and multimedia center, Gifu University, Gifu 501-1193, Japan}
\address[IPMU]{Kavli Institute for the Physics and Mathematics of the Universe (WPI), the University of Tokyo, Kashiwa, Chiba, 277-8582, Japan}
\address[KMS]{Kobayashi-Maskawa Institute for the Origin of Particles and the Universe, 
Nagoya University, Furo-cho, Chikusa-ku, Nagoya, Aichi, 464-8602, Japan.}
\address[KOBE]{Department of Physics, Kobe University, Kobe, Hyogo 657-8501, Japan}
\address[KRISS]{Korea Research Institute of Standards and Science, Daejeon 305-340, South Korea}
\address[MIYA]{Department of Physics, Miyagi University of Education, Sendai, Miyagi 980-0845, Japan}
\address[NAGOYA]{Solar Terrestrial Environment Laboratory, Nagoya University, 
Nagoya, Aichi 464-8602, Japan}
\address[SEJONG]{Department of Physics, Sejong University, Seoul 143-747, South Korea}
\address[TOKAI1]{Department of Physics, Tokai University, Hiratsuka,
  Kanagawa 259-1292, Japan}
\address[TOKAI2]{School of Science and Technology, Tokai University, Hiratsuka,
  Kanagawa 259-1292, Japan}
\address[YNU]{Department of Physics, Faculty of Engineering, Yokohama National University, Yokohama, Kanagawa 240-8501, Japan}

\fntext[tohoku]{Now at Research Center for Neutrino Science, Tohoku University, Sendai 980-8578, Japan}

\begin{abstract}
XMASS, a low-background, large liquid-xenon detector, was used to search for
solar axions that would be produced by bremsstrahlung
and Compton effects in the Sun. With an exposure
of 5.6\,ton days of liquid xenon,
the model-independent limit on the coupling for mass
$\ll$ 1\,keV is $|g_{aee}|< 5.4\times 10^{-11}$ (90\% C.L.),
which is a factor of two stronger than the existing experimental limit.
The bounds on the axion masses for the DFSZ and KSVZ axion models
are 1.9 and 250\,eV, respectively.
In the mass range of 10--40\,keV, this study produced the most stringent
limit, which is better than that previously derived from astrophysical arguments
regarding the Sun to date.
\end{abstract}

\begin{keyword}

Axion, Sun, xenon
\end{keyword}

\end{frontmatter}


\section{Introduction}
The axion is a hypothetical particle invented for
solving the $CP$ problem in strong interactions \cite{PQWW}.
As the initial Peccei--Quinn--Weinberg--Wilczek model
of axions is directly tied to
the electroweak symmetry-breaking scale,
an experimental search was
relatively easy and the model was ruled out early.
However, invisible axion models such as DFSZ \cite{DFSZ}
and KSVZ \cite{KSVZ}, whose symmetry-breaking scale is
separated from the electroweak scale, are still viable.
The DFSZ axions have direct couplings to leptons
whereas the KSVZ axions (hadronic axions) do not have 
tree-level couplings to leptons. In these models, the mass of axions is
\begin{eqnarray}
m_a &=& \frac{\sqrt{z}}{1+z}\frac{f_\pi m_\pi}{f_a}
= \frac{6.0\,{\rm eV}}{f_a/10^6\,{\rm GeV}}, \nonumber
\end{eqnarray}
where $f_a$, $f_\pi$, and $m_\pi$ are
the axion decay constant \cite{Raffelt},
the pion decay constant, and 
pion mass, respectively, and
$z=m_d/m_u\sim 0.56$ is the quark mass ratio.

At present, the search for axions as well as
axion-like particles (ALPs) focuses on couplings
to photons $(g_{a\gamma\gamma})$,
nucleons $(g_{aNN})$
and electrons $(g_{aee})$. There are three types
of searches: (1) laboratory-based experiments in which
sources and detectors are prepared, (2) astrophysical investigations that
examine any significant deviations in the properties of stars
from theoretical predictions due to extra
emission of energy, and (3) using laboratory detectors to look
for axion signals from the Sun or cosmological relics.
Experiments searching for axions have so far produced null results,
but sensitivities continue to improve.

In experimental searches that utilize $g_{a\gamma\gamma}$,
a series of experiments using strong magnets
\cite{Lazarus, sumico, cast} successfully improved sensitivities
by increasing the magnetic field strength
and the conversion length.
The suggestion \cite{Paschos} to use Bragg scattering to improve
sensitivity for solar axions in crystalline detectors
was used in \cite{Avignone1998, Bernabei2001, Avignone2002, CDMS2009}.
Another way to enhance sensitivity is
to exploit resonant absorption on nuclei \cite{Moriyama}.
To date, several experimental results are obtained in this scheme
\cite{Krcmar, Krcmar2001, Krcmar2004, Derbin2007a, Namba,
Derbin2007b, Belli2008, Derbin2009, Derbin2011a, Belli2012, Borexino}.
Significant improvement can be achieved
if the signals can be read out efficiently.
On the other hand, an efficient experimental search with $g_{aee}$
has not been performed.
A pioneering experiment used a Ge detector (710\,g) \cite{Avignone1987}
and a recent search used a Si(Li) detector (1.3\,g) to search for signals
from axions generated by the bremsstrahlung and Compton effect
via the axioelectric effect \cite{Derbin2012}.

The choice of target material strongly affects the reach of
a solar axion experiment using axion coupling to electrons.
Liquid Xe is both dense and has a high atomic numbers \cite{Avignone2009b}.
The XMASS detector,
which uses 835\,kg of liquid xenon in its sensitive volume, is suitable
for this purpose. Its low energy threshold (0.3\,keV) is
also useful as the predicted energy spectrum is very soft
and has a peak at less than 1\,keV for light axions.
Its low background (a few keV$^{-1}$kg$^{-1}$day$^{-1}$) makes it particularly
useful when searching for solar axions.

\section{Expected Signal}
The signals we searched for are produced
by the Compton scattering of photons on electrons $e+\gamma \rightarrow e+a$
and the bremsstrahlung of axions from electrons $e+Z \rightarrow e+a+Z$ in the Sun.
The expected fluxes and spectra are derived as follows.

The solar axion flux produced by Compton scattering
was calculated in \cite{Derbin2011b, Pospelov2008}.
The axion differential flux is expressed as
\begin{eqnarray}
\frac{d\Phi_a^c}{dE_a} &=& \frac{1}{A^2}\int^{R_\odot}_0\!\!\!\!
\int^{\infty}_{E_a}\frac{dN_\gamma}{dE_\gamma}\frac{d\sigma^c}{dE_a}
dE_\gamma N_e(r)r^2dr,
\end{eqnarray}
where $E_a$ is the total energy of the axions,
$A$ is the average distance between the Sun and the Earth,
$R_\odot$ is the radius of the Sun, 
$dN_\gamma/dE_\gamma$ is the blackbody spectrum of photons,
$d\sigma^c/dE_a$ is the cross section for the Compton effect,
and $N_e(r)$ is the electron density at the radius $r$.
Since $m_a$ and $E_\gamma$ is assumed to be much smaller than $m_e$,
the differential cross section is
approximately a product of $\delta(E_a-E_\gamma)$, and
the total cross section \cite{Pospelov2008} is expressed as
\begin{eqnarray}
\sigma^c &=& \alpha 
\frac{g_{aee}^2 E_\gamma^2 v_a}{4m_e^4}\left[\left(1+\frac{v_a^2}{3}\right)
\left(1+\frac{m_a^2}{2E_\gamma^2}\right) -\frac{m_a^2}{E_\gamma^2}
\left(1-\frac{m_a^2}{2E_\gamma^2}\right)\right],
\end{eqnarray}
where $\alpha$ is the fine structure constant,
$m_e$ is the electron mass, $g_{aee}$ is the axion's coupling to electrons \cite{Raffelt}
which is $(1/3)(\cos^2\beta)m_e/f_a$
in the DFSZ axion model \cite{Derbin2012}, and $v_a=(1-m_a^2/E_\gamma^2)^{1/2}$
is the velocity of the outgoing massive axion. $\cot \beta$ is the ratio
of the two Higgs vacuum expectation values of the model \cite{Raffelt}.

The energy spectrum of solar axions produced by the bremsstrahlung effect
was calculated in \cite{Derbin2011b, Zhitnitskii1979}.
The differential energy spectrum is 
\begin{eqnarray}
\frac{d\Phi_a^b}{dE_a} &=& \frac{1}{A^2}\int^{R_\odot}_0\!\!\!\!
\int^{\infty}_{E_a}\frac{dN_e}{dE_e}v_e\frac{d\sigma^b}{dE_a}
dE_e\sum_{Z,A}Z^2N(r)r^2dr,
\end{eqnarray}
where $v_e$ is the velocity of the electrons,
$dN_e/dE_e$ is the energy spectrum of the electrons,
$d\sigma^b/dE_A$ is the cross section for the bremsstrahlung effect,
and $N_{Z,A}(r)$ is the atom density at radius $r$.
The cross section $d\sigma^b/dE_a$ is calculated
by considering the energy conservation of the electron and axion system
\cite{Zhitnitskii1979}.

The temperature, electron density, and atomic density are
given by the standard solar model BP05(OP) \cite{BP05}.
Figure 1 in Ref.\ \cite{Derbin2012} shows
the energy spectra for various masses of axions.
The bremsstrahlung component dominates below 10\,keV,
whereas the Compton contribution dominates at higher energy.

\begin{figure}
\begin{center}
\includegraphics[width=8cm]{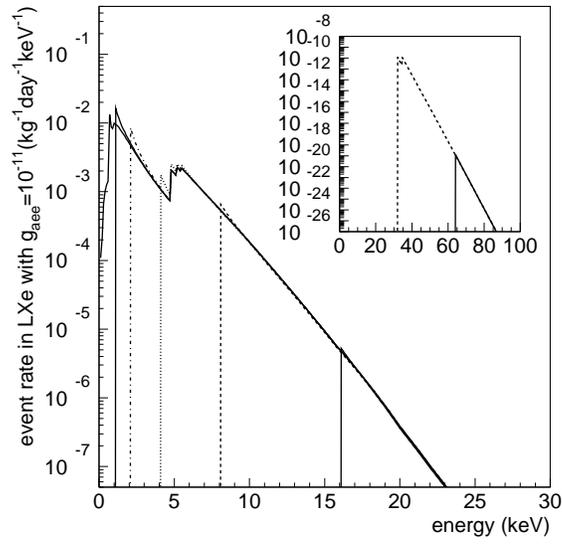}
\end{center}
\caption{Expected energy spectra of events observed using
the liquid-xenon detector. No resolution effects are included.
Different curves
are for axion masses with 0, 1, 2, 4, 8, and 16\,keV. The inset shows
spectra of axion masses with 32 and 64\,keV.
Due to a cross section enhancement for nonrelativistic axions,
an increase at $E \sim m_a$ can be seen. The step around 5\,keV corresponds
to the L-shell absorption edge of the axioelectric effect.
}
\label{xmass:obsspec}
\end{figure}

The expected energy spectrum to be observed with a detector is
\begin{eqnarray}
\frac{dN_{\rm obs}}{dE} &=& \sigma_{ae}(E_a)
\left.\left(\frac{d\Phi^c_a}{dE_a} + \frac{d\Phi^b_a}{dE_a} \right)
\right|_{E_a=E},
\end{eqnarray}
where $\sigma_{ae}(E_a)$ is the cross section
for the axioelectric effect \cite{Pospelov2011}. 
For the cross section, the expression of Eq. (3)
in Ref.\ \cite{Derbin2012} is used for $v_a$
\begin{eqnarray}
\sigma_{ae}(E_a) &=& \sigma_{pe}(E_a)
\frac{g_{aee}^2}{v_a}\frac{3E_a^2}{16\pi\alpha m_e^2}
\left(1-\frac{v_a}{3} \right),
\label{eqn:xsec}
\end{eqnarray}
where $\sigma_{pe}(E_a)$ is the photoelectric cross section
of the detector medium for gamma rays with energy $E_a$.
The photoelectric cross section is available in Ref.\ \cite{XCOM,ATMDATA}.
The predicted energy spectra for a xenon target for various
axion masses are shown in Fig.\ \ref{xmass:obsspec}.

The predicted energy spectra calculated above are used to
generate Monte Carlo simulation samples.
Axion signal samples can be simulated by injecting
gamma rays whose energy is the same as the total energy of 
the incoming axions. This is because (1) there is a relationship
between the cross section of the axioelectric effect
and the photoelectric effect as in Eq. (\ref{eqn:xsec}),
(2) the photoelectric effect
is dominant in this energy range ($<$100\,keV), and
(3) the process after the axioelectric effect is exactly
the same as that for the photoelectric effect.
In the simulation, we considered the nonlinearity of the scintillation
yield for gamma rays, the optical processes of the scintillation
photons in the detector, the photoelectron distributions
and discrimination threshold of photomultipliers,
and the trigger conditions of the data acquisition system.
The detailed description of the simulation
and efficiencies were previously reported \cite{XMASS1,XMASS2}.
After taking into account the reduction efficiency
described in the next section,
the expected energy spectra for various masses of axions
are obtained.

\section{The Data}
The XMASS detector is
a large liquid-xenon detector located underground
(3000\,m water equivalent) at the Kamioka Observatory, Japan. 
It contains an 835-kg liquid-xenon target with a surface of a pentakis-dodecahedron
that is tiled with inward looking photomultiplier tubes (PMT),
630 of which have hexagonal and 12 have round photocathodes.
The PMTs (R-10789, Hamamatsu)
are specially developed for this low-background detector. 
The photoelectron yield at the center of the detector
is evaluated at 14.7\,photoelectrons (p.e.)/keV using an internal $^{57}$Co source.
The positional dependence (maximum 15\%)
of the photoelectron yield caused by the angular acceptance
of PMTs and absorption of scintillation light 
are taken into account in the Monte Carlo simulations. 
Data acquisition is triggered if four or more
PMTs have more than 0.2\,p.e.\ within 200\,ns.
The trigger efficiency around the trigger threshold
was examined by LEDs placed at the detector wall.
The observed behavior was well reproduced by the Monte Carlo simulations.
Signals from each PMT are fed into charge ADCs and TDCs
whose resolution is around 0.05\,p.e.\ and 0.4\,ns, respectively.
The liquid-xenon detector is surrounded by a water Cherenkov
veto counter, which is 10.5\,m in height and 10\,m in diameter.
It is equipped with 72 20-inch PMTs whose signals are
fed into the ADCs and TDCs. Data acquisition is triggered
if eight or more 20-inch PMTs have hits.
The detector is described in detail in Ref.\ \cite{XMASS1}.

The data set used in the solar axion search experiments covers
February 21--27, 2012. A sequence of standard data reduction
is applied to remove events caused by afterpulses
and electronic ringing. The standard reduction consists of a series of
cuts: (1) the event is triggered only by the liquid-xenon detector;
(2) the time difference to the previous event is more than 10\,ms;
(3) the root mean square of the hit timing is less than 100\,ns
and is used to reject events caused by afterpulses of PMTs due to
bright events; and 
(4) the number of PMT hits in the first 20\,ns
divided by the total number of hits is less than 0.6 for 
events in which the number of photoelectrons is less than 200.
The fourth cut was applied to remove Cherenkov events
originated from $^{40}$K in photocathodes (Cherenkov cut).
The energy threshold of this analysis is low (0.3\,keV)
because of our exceptional photoelectron yield,
which is the largest among current low-background detectors.
A more detailed description of the reduction can be found in
Ref.\ \cite{XMASS2}.

\begin{figure}
\begin{center}
\includegraphics[width=8cm]{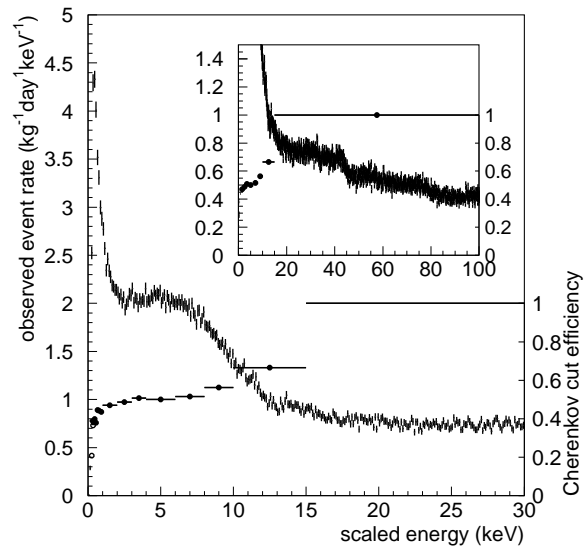}
\end{center}
\caption{Observed energy spectra. The horizontal
axis shows the ``scaled energy'' calculated
by dividing the number of photoelectrons by the photoelectron yield at
the center of the detector, 14.7\,p.e./keV. Error bars
are statistical only. In this figure we also show the efficiencies
for the Cherenkov cut (closed circles with
horizontal bars for the applicable range; 1 for 100\%) and 
for the combination of all our cuts (open circles).
Only at the trigger threshold is the overall efficiency not dominated
by the Cherenkov cut efficiency.
The inset shows the same quantities for
energies extending up to 100\,keV.}
\label{xmass:obs}
\end{figure}

Figure \ref{xmass:obs} shows the observed energy spectra.
The total livetime is 6.7\,days
after considering the dead time caused by the cut (2).
The effect of trigger cut (1) is visible below 0.4\,keV 
as shown in Fig.\ 3 in Ref.\ \cite{XMASS2} and is considered
in our Monte Carlo simulations. The same samples show that the cut
(3) has negligible effect on the signals.
The signal efficiency due to the Cherenkov cut, which is drawn in the same figure,
was conservatively evaluated using low-energy gamma-ray sources
such as $^{55}$Fe and $^{241}$Am sources at various positions.
Because the efficiency weakly depends on the radial position
of the events and gradually decreases outward,
the efficiency adopted in the analysis was
mostly evaluated at a radius of 40\,cm
where 93\% of the mass was contained inside.
The Monte Carlo samples were compared with the observed
energy spectra after weighting this efficiency.

\begin{figure}
\begin{center}
\includegraphics[width=8cm]{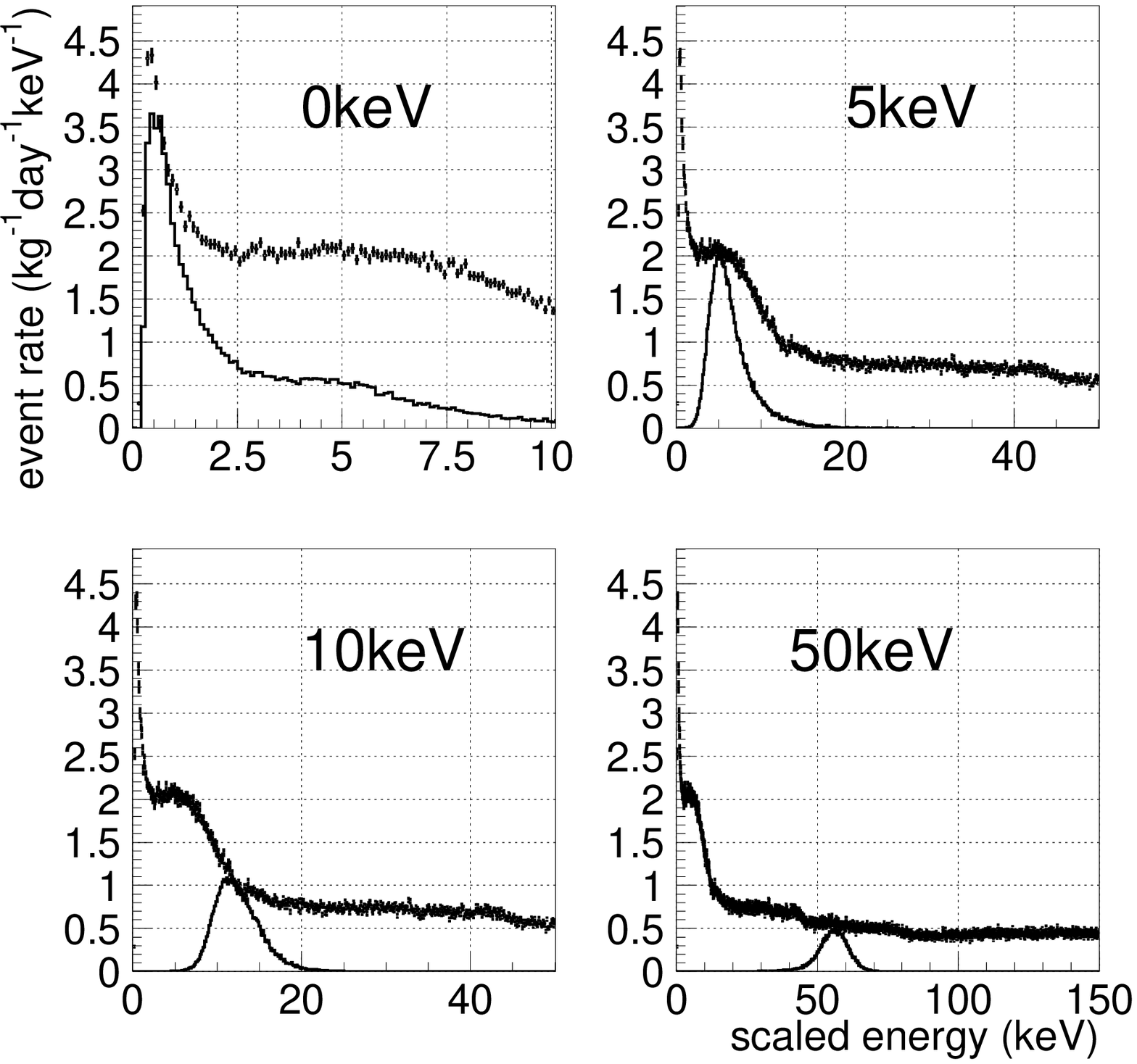}
\end{center}
\caption{Comparison between the observed data (points with error bars) and
expected spectrum (solid histogram) for axion masses of 0, 5, 10, and 50\,keV.
The solid histograms are scaled to the maximum coupling
allowed at 90\% C.L.}
\label{xmass:comparison}
\end{figure}

\section{Limit on $g_{aee}$}
\begin{figure}
\begin{center}
\includegraphics[width=8cm]{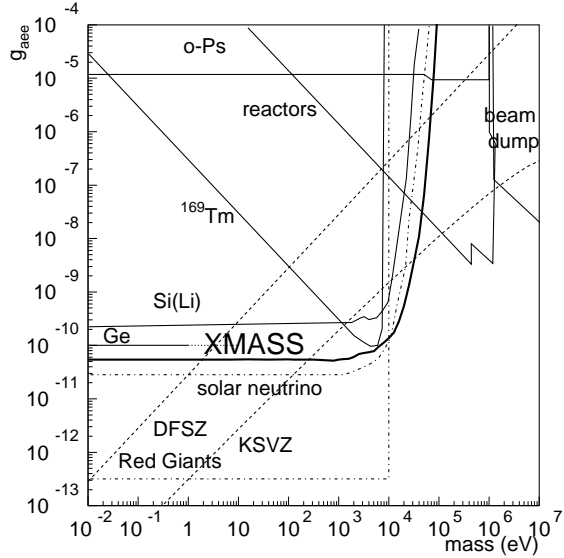}
\end{center}
\caption{Limits on $g_{aee}$. The thick solid line
shows the limit obtained in this study.
The other solid lines are limits obtained by laboratory experiments:
Ge \cite{Avignone1987}, Si(Li), $^{169}$Tm, reactors, $o$-Ps, and
beam-dump experiments (see \protect\cite{Derbin2012} and references therein.)
The dash--dotted lines show astrophysical limits from 
red giant stars \protect\cite{Raffelt}
and the solar neutrino flux \protect\cite{Gondolo2009}.
The dashed lines are theoretical predictions for the
DFSZ $(\cos^2\beta=1)$ and KSVZ $(E/N=8/3)$ models.
This study gives a stronger constraint
by a factor of two over previous direct experimental limits
for axion mass $\ll$1\,keV, and
the best constraint absolute between 10 and 40\,keV.
}
\label{xmass:limit}
\end{figure}

The observed spectra do not have any prominent features
to identify axion signals with respect to the
background. Instead, strong constraints on $g_{aee}$ can be obtained
from the observed event rate in the relevant energy range.
In order to set a conservative upper limit on the axion-electron 
coupling constant $g_{aee}$, the coupling is adjusted until
the expected event rate in XMASS does not exceed the one observed
in any energy bin above 0.3\,keV.
Figure \ref{xmass:comparison} shows the expected energy spectra
with the coupling constants obtained by the procedure above.
Figure \ref{xmass:limit} shows the summary of the bounds of $g_{aee}$.
For small axion masses,
a $g_{aee}$ value of $5.4\times 10^{-11}$ is obtained. This is 
the best direct experimental limit to date and is close
to that derived from astrophysical considerations based on
measured solar neutrino fluxes:  $g_{aee} = 2.8\times 10^{-11}$ \cite{Gondolo2009}.
For axion masses $>$ 10\,keV the energetics in the Sun
are no longer sufficient to effectively produce such axions.
A systematic uncertainty inherent to our method of comparing 
bin contents arises from the specific choice of binning.
This and systematic uncertainties for energy scale including
energy threshold, Cherenkov cut efficiency,
and energy resolution are evaluated to be 2\%, 1\%, 2\%,
and 1\%, respectively. The total systematic error, 3\%,
is obtained by summing these contributions in quadrature,
and the limit in Fig.\ \ref{xmass:limit} (90\% C.L.)
takes this error into account.

The calculated limit depends on the 
interaction processes considered in our detector
as well as the processes considered for solar axion production in the Sun.
Processes such as the inverse Primakoff effect and
nuclear absorption on the detection side,
and the Primakoff effect and nuclear deexcitation
on the production side can be neglected because the constraints
on $g_{g\gamma\gamma}$ and $g_{aNN}$ are tight.
A possible additional contribution caused by $g_{aee}$ on the detection side
is the inverse Compton effect. This can be neglected
because of its small cross section \cite{Bernabei2006}.
On the production side, there are other known contributions
such as electron-electron bremsstrahlung \cite{Raffelt1986}
and the axio-recombination effect \cite{axiorecombination}.
However, the expected fluxes for these processes are only known
in the limit of massless axions. For this reason and in order
to directly compare our results with the most relevant
previously published ones we restrict the production processes
we consider to the electron-nuclei bremsstrahlung and the Compton
effect. As omitting production mechanisms lowers the flux estimate, 
all the limits thus derived will have to be considered conservative.

The nature of the events surviving the analysis cuts is also of interest.
According to our study on these events, 
most of them originate on the inner surface of the detector \cite{IDM2012YS}.
These events are attributed to radioactive contamination in the aluminum
seal of the PMT entrance windows, $^{14}$C decays in the GORE-TEX$^{\textregistered}$
sheets between the PMTs and the copper support
structure, and light leaking from gaps in between the triangular
elements of this support structure.

\section{Conclusion}
In summary, solar axions produced through axion-electron coupling were
searched for in XMASS, a large liquid-xenon detector.
The energy threshold is low (0.3\,keV)
because of our exceptional photoelectron yield,
which is the largest among current low-background detectors.
As our observed spectrum does not show any indications
of axion signals, we derive constraints on the $g_{aee}$ coupling.
Our limit on $g_{aee}$ for axions with mass much smaller
than 1\,keV is $5.4\times 10^{-11}$.
The bounds on the axion masses for the DFSZ and KSVZ axion models
are 1.9 and 250\,eV, respectively.
For axion masses between 10 and 40\,keV, our new limits are the most stringent
that are currently available.

\section*{Acknowledgements}
We gratefully acknowledge the cooperation of Kamioka Mining
and Smelting Company. 
This work was supported by the Japanese Ministry of Education,
Culture, Sports, Science and Technology, Grant-in-Aid
for Scientific Research, and partially
by the National Research Foundation of Korea Grant funded
by the Korean Government (NRF-2011-220-C00006).
We thank J.~Redondo for useful discussion.





\bibliographystyle{model1-num-names}
\bibliography{<your-bib-database>}



\end{document}